\documentclass[longauth]{aa}
\usepackage{graphicx}
\usepackage{txfonts}
\usepackage{xcolor}
\usepackage{hyperref}
\hypersetup{colorlinks, linkcolor=blue, citecolor=blue, urlcolor=blue}

\newcommand{\pivec}{\mbox{\boldmath $\pi$}}
\newcommand{\muvec}{\mbox{\boldmath $\mu$}}

\newcommand{\te}{t_{\rm E}}
\newcommand{\thetae}{\theta_{\rm E}}

\newcommand{\pie}{\pi_{\rm E}}
\newcommand{\pien}{\pi_{{\rm E},N}}
\newcommand{\piee}{\pi_{{\rm E},E}}
\newcommand{\dl}{D_{\rm L}}

\definecolor{brown}{rgb}{0.59, 0.29, 0.0}
\definecolor{darkgreen}{rgb}{0.0, 0.42, 0.24}
\definecolor{darkblue}{rgb}{0.01, 0.31, 0.59}
\definecolor{darkblue}{rgb}{0.0, 0.25, 0.42}
\definecolor{blue}{rgb}{0.0,0.0,1.0}
\definecolor{green}{rgb}{0.0,1.0,0.0}

\begin{document}

\title{Brown-dwarf companions in microlensing binaries detected during the 2016--2018 seasons}

\author{
     Cheongho Han\inst{01}
\and Yoon-Hyun Ryu\inst{02}
\and In-Gu Shin\inst{01}
\and Youn Kil Jung\inst{02}
\and Doeon Kim\inst{01}
\and Yuki Hirao\inst{03}
\and Valerio Bozza\inst{04,05}
\and Michael D. Albrow\inst{06}
\and Weicheng Zang\inst{07}
\and Andrzej Udalski\inst{08}
\and Ian~A.~Bond\inst{09}
\\
(Leading authors)\\
     Sun-Ju Chung\inst{02}
\and Andrew Gould\inst{10,11}
\and Kyu-Ha Hwang\inst{02}
\and Yossi Shvartzvald\inst{12}
\and Hongjing Yang\inst{07}
\and Sang-Mok Cha\inst{02,13}
\and Dong-Jin Kim\inst{02}
\and Hyoun-Woo Kim\inst{02}
\and Seung-Lee Kim\inst{02}
\and Chung-Uk Lee\inst{02}
\and Dong-Joo Lee\inst{02}
\and Jennifer~C.~Yee\inst{15}
\and Yongseok Lee\inst{02,14}
\and Byeong-Gon Park\inst{02,14}
\and Richard W. Pogge\inst{11}
\\
(The KMTNet collaboration)\\
     Przemek Mr{\'o}z\inst{08}
\and Micha{\l} K. Szyma{\'n}ski\inst{08}
\and Jan Skowron\inst{08}
\and Radek Poleski\inst{08}
\and Igor Soszy{\'n}ski\inst{08}
\and Pawe{\l} Pietrukowicz\inst{08}
\and Szymon Koz{\l}owski\inst{08}
\and Krzysztof Ulaczyk\inst{16}
\and Krzysztof A. Rybicki\inst{08,12}
\and Patryk Iwanek\inst{08}
\and Marcin Wrona\inst{08}
\\
(The OGLE Collaboration)\\
     Fumio Abe\inst{17}
\and Richard Barry\inst{18}
\and David P. Bennett\inst{18,19}
\and Aparna Bhattacharya\inst{18,19}
\and Hirosame Fujii\inst{17}
\and Akihiko~Fukui\inst{20,21}
\and Stela Ishitani Silva\inst{18,22}
\and Rintaro Kirikawa\inst{03}
\and Iona Kondo\inst{03}
\and Naoki Koshimoto\inst{23}
\and Yutaka Matsubara\inst{17}
\and Sho~Matsumoto\inst{03}
\and Shota Miyazaki\inst{03}
\and Yasushi Muraki\inst{17}
\and Arisa Okamura\inst{03}
\and Greg Olmschenk\inst{18}
\and Cl{\'e}ment Ranc\inst{24}
\and Nicholas J. Rattenbury\inst{25}
\and Yuki Satoh\inst{03}
\and Takahiro Sumi\inst{03}
\and Daisuke Suzuki\inst{03}
\and Taiga Toda\inst{03}
\and Paul~J.~Tristram\inst{26}
\and Aikaterini Vandorou\inst{18,19}
\and Hibiki Yama\inst{03}
\and Yoshitaka Itow\inst{17}
\\
(The MOA Collaboration)\\
}

\institute{
       Department of Physics, Chungbuk National University, Cheongju 28644, Republic of Korea,                                                           
\and  Korea Astronomy and Space Science Institute, Daejon 34055, Republic of Korea                                                                       
\and  Department of Earth and Space Science, Graduate School of Science, Osaka University, Toyonaka, Osaka 560-0043, Japan                               
\and  Dipartimento di Fisica "E. R. Caianiello", Universit\'a di Salerno, Via Giovanni Paolo II, 84084 Fisciano (SA), Italy                              
\and  Istituto Nazionale di Fisica Nucleare, Sezione di Napoli, Via Cintia, 80126 Napoli, Italy                                                          
\and  University of Canterbury, Department of Physics and Astronomy, Private Bag 4800, Christchurch 8020, New Zealand                                    
\and  Department of Astronomy, Tsinghua University, Beijing 100084, China                                                                                
\and  Astronomical Observatory, University of Warsaw, Al.~Ujazdowskie~4, 00-478~Warszawa, Poland                                                         
\and  Institute of Natural and Mathematical Science, Massey University, Auckland 0745, New Zealand                                                       
\and  Max-Planck-Institute for Astronomy, K\"{o}nigstuhl 17, 69117 Heidelberg, Germany                                                                   
\and  Department of Astronomy, Ohio State University, 140 W. 18th Ave., Columbus, OH 43210, USA                                                          
\and  Department of Particle Physics and Astrophysics, Weizmann Institute of Science, Rehovot 76100, Israel                                              
\and  School of Space Research, Kyung Hee University, Yongin, Kyeonggi 17104, Republic of Korea                                                          
\and  Korea University of Science and Technology, Korea, (UST), 217 Gajeong-ro, Yuseong-gu, Daejeon, 34113, Republic of Korea                            
\and  Center for Astrophysics $|$ Harvard \& Smithsonian, 60 Garden St., Cambridge, MA 02138, USA                                                        
\and  Department of Physics, University of Warwick, Gibbet Hill Road, Coventry, CV4~7AL,~UK                                                              
\and  Institute for Space-Earth Environmental Research, Nagoya University, Nagoya 464-8601, Japan                                                        
\and  Code 667, NASA Goddard Space Flight Center, Greenbelt, MD 20771, USA                                                                               
\and  Department of Astronomy, University of Maryland, College Park, MD 20742, USA                                                                       
\and  Department of Earth and Planetary Science, Graduate School of Science, The University of Tokyo, 7-3-1 Hongo, Bunkyo-ku, Tokyo 113-0033, Japan      
\and  Instituto de Astrof{\'i}sica de Canarias, V{\'i}a L{\'a}ctea s/n, E-38205 La Laguna, Tenerife, Spain                                               
\and  Department of Physics, The Catholic University of America, Washington, DC 20064, USA                                                               
\and  Department of Astronomy, Graduate School of Science, The University of Tokyo, 7-3-1 Hongo, Bunkyo-ku, Tokyo 113-0033, Japan                        
\and  Zentrum f{\"u}r Astronomie der Universit{\"a}t Heidelberg, Astronomisches Rechen-Institut, M{\"o}nchhofstr.\ 12-14, 69120 Heidelberg, Germany      
\and  Department of Physics, University of Auckland, Private Bag 92019, Auckland, New Zealand                                                            
\and  University of Canterbury Mt. John Observatory, P.O. Box 56, Lake Tekapo 8770, New Zealand                                                          
}
%

\abstract
{}
{
With the aim of finding microlensing binaries containing brown-dwarf (BD) companions, we investigate 
the microlensing survey data collected during the 2016--2018 seasons.  
}
{
For this purpose, we first conducted modeling of lensing events with light curves exhibiting 
anomaly features that are likely to be produced by binary lenses.  We then sorted out BD-companion 
binary-lens events by applying the criterion that the companion-to-primary mass ratio is $q \lesssim 0.1$.  
From this procedure, we identify 6 binaries with candidate BD companions, including OGLE-2016-BLG-0890L,
MOA-2017-BLG-477L, OGLE-2017-BLG-0614L, KMT-2018-BLG-0357L, OGLE-2018-BLG-1489L, and OGLE-2018-BLG-0360L.
}
{
We estimate the masses of the binary companions by conducting Bayesian analyses using the 
observables of the individual lensing events.  According to the Bayesian estimation of the 
lens masses, the probabilities for the lens companions of the events OGLE-2016-BLG-0890, 
OGLE-2017-BLG-0614, OGLE-2018-BLG-1489, and OGLE-2018-BLG-0360 to be in the BD mass regime 
are very high with $P_{\rm BD}> 80\%$.  For MOA-2017-BLG-477 and KMT-2018-BLG-0357, the 
probabilities are relatively low with $P_{\rm BD}=61\%$ and 69\%, respectively.
}
{}


\maketitle

\section{Introduction}\label{sec:one}

One important scientific feature of microlensing is its capability of detecting faint or dark
objects. For this reason, a microlensing experiment was originally proposed to search for dark
matter in the form of massive compact halo objects lying in the Galactic halo \citep{Paczynski1986}.
Since the completion of the first-generation experiments conducted for this purpose, for example, 
MACHO \citep{Alcock1996}, EROS \citep{Aubourg1995}, and OGLE \citep{Udalski1993}, the application
of microlensing was expanded to looking for faint binary companions to stars \citep{Mao1991}, 
including planets and brown dwarfs (BDs). At the time of writing this paper, 173 microlensing 
planets were reported according to the Extrasolar Planets Encyclopaedia (http://exoplanet.eu/). 
The list of microlensing BDs is given in Table~1 of \citet{Chung2019}, which includes 18 BDs, in 
addition to which there are 9 BDs or candidates that have been reported since that time 
\citep{Shvartzvald2019, Jung2018, Miyazaki2018, Han2020a, Han2020b, Herald2022}.

\begin{table*}[t]
\small
\caption{ID references, alert dates, and coordinates of lensing events \label{table:one}}
\begin{tabular}{lllll}
\hline\hline
\multicolumn{1}{c}{KMTNet}                      &
\multicolumn{1}{c}{OGLE}                        &
\multicolumn{1}{c}{MOA}                         &
\multicolumn{1}{l}{(RA, DEC)$_{\rm J2000}$}     \\
\multicolumn{1}{c}{}                            &
\multicolumn{1}{c}{}                            &
\multicolumn{1}{c}{}                            &
\multicolumn{1}{l}{$(l,b)$}                     \\
\hline
KMT-2016-BLG-0793         & {\bf OGLE-2016-BLG-0890} (0934)  &                           &  (17:30:25.69, -29:50:48.98) \\
(postseason)              & (2016-05-18)                     &                           &  $(-2^\circ\hskip-2pt .540, 2^\circ\hskip-2pt .308)$ \\
\hline                                                                                     
KMT-2017-BLG-1757         &                                  &  {\bf MOA-2017-BLG-477}   &  (18:05:50.00, -27:04:38.50) \\
(postseason)              &                                  &  (2017-09-15)             &  $(3^\circ\hskip-2pt .854, -2^\circ\hskip-2pt .918)$ \\
\hline                                                                                     
KMT-2017-BLG-2209         & {\bf OGLE-2017-BLG-0614}         &                           &  (17:26:08.08, -30:17:46.14) \\
(postseason)              & (2017-04-23)                     &                           &  $(-3^\circ\hskip-2pt .430, 2^\circ\hskip-2pt .833)$ \\
\hline                                                                                     
{\bf KMT-2018-BLG-0357}   &                                  &                           &  (17:44:12.20, -33:36:23.18) \\
(2018-06-30)              &                                  &                           &  $(-4^\circ\hskip-2pt .143, 2^\circ\hskip-2pt .180)$, \\
\hline                                                                                     
KMT-2018-BLG-1534         & {\bf OGLE-2018-BLG-1489}         &                           &  (17:45:46.60, -23:57:43.85) \\
(postseason)              & (2018-08-12)                     &                           &  $(4^\circ\hskip-2pt .267, 2^\circ\hskip-2pt .559)$ \\
\hline                                                                                     
KMT-2018-BLG-2014         & {\bf OGLE-2018-BLG-0360}         &  MOA-2018-116             &  (17:52:01.26, -31:08:54.71) \\
(postseason)              & (2018-03-15)                     &  (2018-04-22)             &  $(1^\circ\hskip-2pt .183, -2^\circ\hskip-2pt.326)$ \\
\hline
\end{tabular}
\end{table*}

The microlensing signature of a planet, with a planet-to-host mass ratio of order $10^{-3}$ or 
less, can be, in most cases, readily identified from its characteristic signature of a short-term 
anomaly to the lensing light curve produced by the host of the planet \citep{Gould1992a}.  By 
contrast, an immediate identification of a BD companion belonging to a binary lens is generally 
much more difficult because the lensing light curve produced by a binary containing a BD companion, 
with a mass ratio between the BD companion to its primary of order $10^{-2}$, is not much different 
from those produced by binaries composed of roughly equal mass components, and thus it does not 
usually exhibit a characteristic pattern that would enable one to immediately identify the 
existence of a BD companion.\footnote{A lensing event produced by a giant planet lying at around 
the Einstein ring of the host also result in a lensing light curve with a planet signal that 
significantly deviates from a short-term anomaly \citep{Han2021b}.} This implies that identifying 
binaries with BD companions requires modeling all the lensing light curves of more than a hundred 
binary lensing events that are being annually detected by the current lensing surveys. As will be 
discussed below, binary-lens modeling requires heavy computations not only because of the large 
number of parameters required to be included in the modeling but also because of the need to 
employ numerical methods.

With the aim of finding binaries containing BD companions, we investigated the microlensing survey 
data. In this paper, we report 6 binaries with candidate BD companions found from the investigation 
of the three years of microlensing data obtained during the seasons from 2016 to 2018, including 
OGLE-2016-BLG-0890L, MOA-2017-BLG-477L, OGLE-2017-BLG-0614L, KMT-2018-BLG-0357L, OGLE-2018-BLG-1489L, 
and OGLE-2018-BLG-0360L.

For the presentation of the work, we arrange the paper according to the following organization. 
In Sect.~\ref{sec:two}, we describe the data used in the analysis, the instruments used for the 
acquisition of the data, and the procedure of data reduction.  In Sect.~\ref{sec:three}, we mention 
the procedure of modeling lensing light curves and the criteria applied to sort out BD-companion 
binary-lens events.  In the following subsections (\ref{sec:three-one}--\ref{sec:three-six}), we 
explain the details of the modeling conducted for the individual lensing events, and present the 
lensing parameters and configurations of the lens systems.  In Sect.~\ref{sec:four}, we specify the 
source stars, measure their angular radii, and estimate the Einstein radii of the individual events.  
In Sect.~\ref{sec:five}, we estimate the physical parameters of the lenses, including the masses of 
the binary components and distances to the lens systems.  In Sect.~\ref{sec:six}, we summarize 
results found from the analyses and conclude.

\section{Observations and data}\label{sec:two}

For the searches of BD events, we first investigated the data of the Korea Microlensing Telescope
Network \citep[KMTNet:][]{Kim2016} survey collected during the first three years of its full 
operation from 2016 to 2018.  During these seasons, 2588, 2817, and 2781 lensing events were 
found by the KMTNet survey in the 2016, 2017, and 2018 seasons, respectively.  Among these events, 
we conducted systematic analyses of anomalous events, for which lensing light curves exhibited 
deviations from the form of single-lens single-source (1L1S) events \citep{Paczynski1986}.  Analyses 
conducted for the individual events vary depending on the nature of the anomalies, for example, 
planetary \citep{Han2020c}, binary-lens \citep{Han2019}, binary-source \citep{Jung2017}, triple-lens 
\citep{Han2022}, binary-lens binary-source \citep{Han2021a} modeling, etc., and details of the 
analyses for different types of anomalies are described in the cited references.  For candidate BD 
events found from this investigation, we conduct detailed analyses using improved data processed 
from optimized photometry of the events.  We then check whether the events were additionally 
observed by the two other working lensing surveys of the Optical Gravitational Lensing Experiment 
\citep[OGLE:][]{Udalski2015} and the Microlensing Observations in Astrophysics survey 
\citep[MOA:][]{Bond2001} in order to include these additional data in the analyses.

From the investigation, we found 6 candidate BD binary events, including
KMT-2016-BLG-0793/OGLE-2016-BLG-0934, KMT-2017-BLG-1757/MOA-2017-BLG-477,
KMT-2018-BLG-1534/OGLE-2018-BLG-1489, KMT-2018-BLG-0357,
KMT-2018-BLG-1534/OGLE-2018-BLG-1489, and
KMT-2018-BLG-2014/OGLE-2018-BLG-0360/MOA-2018-BLG-116.
Among them, four events were observed by two surveys, one was observed by all the three 
surveys, and one was observed solely by the KMTNet survey. In Table~\ref{table:one}, we 
summarize the ID references of the events assigned by the individual survey groups together 
with the alert dates and coordinates.  This paper is the first release of BD events found 
from the systematic investigation of the KMTNet data collected during 2016 -- 2018 seasons, 
and we plan to search for more BD events by investigating the data of the subsequent seasons.

For the events observed by multiple surveys, we hereafter use the ID 
references of the first discovery surveys, marked in bold font in Table~\ref{table:one}, for the 
designation of the events. The notation ``postseason'' for the KMTNet events indicates that the 
events were found from the post-season investigation of the data \citep{Kim2018a}.  We note that 
OGLE and MOA lensing events were found in real time with the progress of the events during the 
2016 -- 2018 seasons, but the real-time alert by the KMTNet survey, the AlertFinder algorithm 
\citep{Kim2018b}, has been operated since the 2018 season.  There are two ID references for 
OGLE-2016-BLG-0890 (the other being OGLE-2016-BLG-0934) because the source  of the event was 
located in two OGLE fields.  In this case, we use both data sets.

Observations by the KMTNet survey were carried out by employing three identical telescopes, each
of which has a 1.6~m aperture. The KMTNet telescopes are globally distributed in three continents
of the Southern Hemisphere, and the sites of the individual telescopes are the Siding Spring 
Observatory in Australia (KMTA), the Cerro Tololo Interamerican Observatory in Chile (KMTC), and 
the South African Astronomical Observatory in South Africa (KMTS). The telescopes used by the OGLE 
and MOA surveys are located at the Las Campanas Observatory in Chile and the Mt.~John Observatory 
in New Zealand, respectively, and the individual telescopes have 1.3~m and 1.8~m apertures. 
The KMTNet, OGLE, and MOA telescopes are equipped with cameras yielding 4~deg$^2$, 1.4~deg$^2$, and 
2.2~deg$^2$ fields of view, respectively.  The main observations by the KMTNet and OGLE surveys 
were done in the $I$ band, while MOA observations were done in the customized MOA-$R$ band. For all 
surveys, a fraction of images were obtained in the $V$ band for the color measurements of source 
stars.  The reductions of data were done using the photometry pipelines of the individual survey 
groups: \citet{Albrow2009} for KMTNet, \citet{Udalski2003} for OGLE, and \citet{Bond2001} for 
MOA. For each data set, the error bars from the photometry pipelines were readjusted to make the 
data consistent with the scatter of data and so that the $\chi^2$ per degree of freedom becomes unity 
following the \citet{Yee2012} routine.

\section{Procedures of event selection and modeling}\label{sec:three}

We search for BDs belonging to binary lenses rather than single-mass BDs for two major reasons.
First, the lensing parameter of the companion-to-primary mass ratio, $q$, can be securely
measured for general binary-lens events, and thus it is possible to pick out BD candidates from
the measured mass ratios. Considering that typical Galactic lensing events are produced by low-mass 
stars 
\citep{Han2003},
companions of binary lenses with mass ratios $q\lesssim 10^{-1}$ are very likely to be BDs. 
Second, anomalies in binary-lens events often involve caustics, and thus it is possible 
to measure an extra observable of the angular Einstein radius $\thetae$, which is difficult to be
measured for general single-lens events, but see \citet{Gould2022} for a systematic study.  While 
the event time scale, $\te$, which is the basic lensing parameter measurable for both single and 
binary lens events, is related to the three physical lens parameters of the mass, $M$, and distance 
to the lens, $\dl$, and the relative lens-source proper motion, $\mu$, the Einstein radius is related 
to the two parameters of $M$ and $\dl$.  Therefore, the mass of the companion can be more tightly 
constrained with the additional measurement of $\thetae$.

The binary-lens (2L1S) modeling of each lensing event is conducted following the common procedure 
described below.  In the modeling, we search for a set of lensing parameters (solution) describing 
the observed lensing light curve. Under the assumption of a rectilinear lens-source relative motion 
(standard model), a binary-lensing light curve is described by 7 lensing parameters. The first three 
parameters $(t_0, u_0, \te)$ depict the approach of the source to the lens, and the individual 
parameters denote the time of the closest lens-source approach, the separation at that time (impact 
parameter), and the event time scale, respectively. The impact parameter is scaled to $\thetae$.  
Three other parameters $(s, q, \alpha)$ describe the binarity of the lens, and they represent the 
projected separation (normalized to $\thetae$) and mass ratio between the lens components, and the 
angle between the relative lens-source motion and the axis connecting the lens components (source 
trajectory angle).  The last parameter $\rho$ (normalized source radius), which is defined as 
the ratio of the angular source radius $\theta_*$ to $\thetae$, is included in modeling, because a 
binary-lensing light curve usually exhibits anomalies resulting from caustic crossings or approaches, 
during which the light curve is affected by finite-source effects \citep{Bennett1996}.

Caustics represent source positions at which lensing magnifications of a point source becomes
infinity. Binary caustics exist in three types of topology, referred to as ``close'', 
``intermediate'', and ``wide'' binaries \citep{Erdl1993, Dominik1999}. In the regime of the 
wide binary ($s\gg 1$), two sets of caustic form near the individual lens components. In the 
close-binary regime ($s\ll 1$), there exist three caustic sets, for which one lies near the 
primary and the other two lie away from the binary axis on the opposite side of the lens 
companion.  In the intermediate regime, multiple sets of caustics merge to form a single large 
caustic, which is often referred to as a ``resonant'' caustic. See Figure~1 of \citet{Cassan2008} 
for the caustic topologies in the three regimes of binary lenses.

Besides the basic parameters, binary-lens modeling occasionally requires one to include extra
parameters for the description of higher-order effects in lensing light curves.  One such  
higher-order effect is caused by the deviation of the source motion from a rectilinear one 
induced by the orbital motion of Earth: microlens-parallax effect \citep{Gould1992b}. Another 
higher-order effect is caused by the variation of the lens position induced by the orbital 
motion of the binary lens: lens-orbital effect \citep{Albrow2000}.  Considering these higher-order 
effects in modeling requires inclusion of additional parameters, which are $(\pien, \piee)$ for 
the microlens-parallax effect and $(ds/dt, d\alpha/dt)$ for the lens-orbital effects. The parameters 
$(\pien, \piee)$ denote components of the microlens-parallax vector $\pivec_{\rm E}$ projected onto 
the sky along the north and east directions, respectively, and the parameters $(ds/dt, d\alpha/dt)$ 
represent the change rates of the binary separation and source trajectory angle, respectively. 
We note that the 2-parameter description of the lens orbital motion is a local approximation 
of a more complete Keplerian model.  The microlens-parallax vector is related to the relative 
lens-source parallax $\pi_{\rm rel}={\rm AU} (D_{\rm L}^{-1}-D_{\rm S}^{-1})$ and the relative 
lens-source proper motion $\muvec$ by $\pivec_{\rm E}=(\pi_{\rm rel}/\thetae)(\muvec/\mu)$.

The binary-lens modeling is carried out in two steps. In the first step, we search for the binary
parameters $(s, q)$ using a grid approach, while the other parameters are found using a downhill
approach. We use the Markov Chain Monte Carlo (MCMC) algorithm for the downhill approach.  This 
first-step procedure yields a $\Delta\chi^2$ map on the $s$--$q$ parameters plane, and we identify 
local solutions on the map, including those resulting from various types of degeneracy, if they 
exist.  In the second step, we polish the individual local solutions by letting all parameters, 
including $s$ and $q$, vary. We present multiple solutions if degeneracies among different solutions 
are severe. We also check higher-order effects for well-covered lensing events with long time scales.  
For some anomalous events with no obvious caustic-crossing features, we additionally check whether 
the observed anomalies can be explained by a binary-source interpretation \citep{Gaudi1998}.  In the 
subsequent subsections, we present the analyses of the individual events.

\subsection{OGLE-2016-BLG-0890}\label{sec:three-one}

The source of the lensing event OGLE-2016-BLG-0890 lies toward the Galactic bulge field with the
equatorial coordinates $({\rm RA}, {\rm DEC})_{\rm J2000}=($17:30:25.69, -29:50:48.98), which 
correspond to the Galactic coordinates $(l, b)=(-2^\circ\hskip-2pt .540, 2^\circ\hskip-2pt .308)$. 
The baseline magnitude of the source is $I_{\rm base}=16.25$. The event was first found by the OGLE 
survey on 2016 May 18 (${\rm HJD}^\prime\equiv {\rm HJD}-2450000 \sim 7516$), at which the source 
became brighter than the baseline by $\sim 0.16$~mag.  The event was also in the footprint of the 
KMTNet survey, and it was identified from the post-season investigation \citep{Kim2018a} and 
designated as KMT-2016-BLG-0793.

\begin{figure}[t]
\includegraphics[width=\columnwidth]{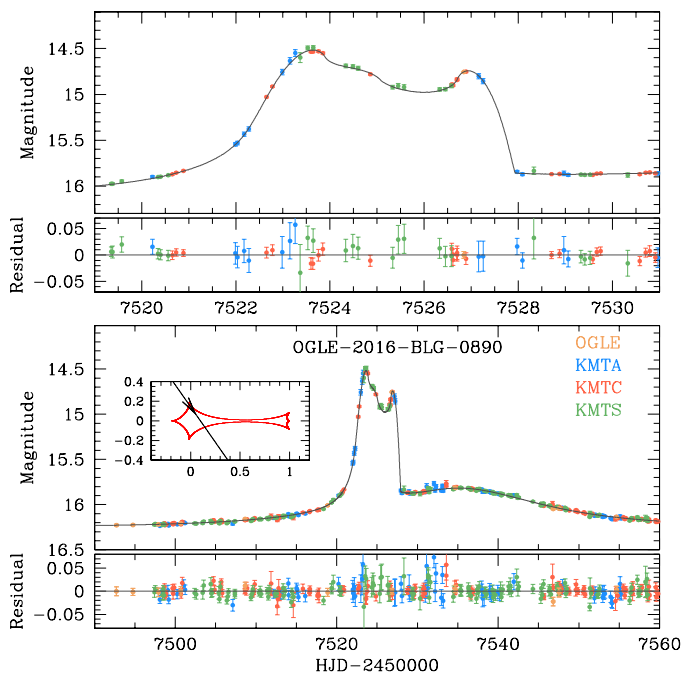}
\caption{
Light curve and model of OGLE-2016-BLG-0890. The upper panel shows the
enlargement of the major anomaly region and the residual from the model. The inset in the lower
panel shows the lens-system configuration, in which the line with an arrow represents the source
trajectory and the red figure is the caustic.
}
\label{fig:one}
\end{figure}

The lensing light curve constructed from the combination of the OGLE and KMTNet data is presented
in Figure~\ref{fig:one}. It shows clear features of caustic crossings at ${\rm HJD}^\prime=7522.35$ 
and 7527.00, which correspond to the times of the caustic entrance and exit, respectively. Both 
caustic crossings were resolved by the data obtained from the combination of the KMTNet observations 
conducted with a 2.5 hour cadence. The light curve in the region between the two caustic-crossing 
features exhibits deviations from a typical U-shape pattern, and this suggests that the source 
passed along a fold of the caustic.

\begin{table}[t]
\small
\caption{Model of OGLE-2016-BLG-0890\label{table:two}}
\begin{tabular*}{\columnwidth}{@{\extracolsep{\fill}}lcccc}
\hline\hline
\multicolumn{1}{c}{Parameter}    &
\multicolumn{1}{c}{Value }       \\
\hline
$\chi^2$/dof            &   $723.97/717           $      \\
$t_0$ (HJD$^\prime$)    &   $7525.298 \pm 0.012   $      \\
$u_0$                   &   $0.0817 \pm 0.0010    $      \\
$\te$ (days)            &   $15.00 \pm 0.06       $      \\
$s$                     &   $1.594 \pm 0.002      $      \\
$q$                     &   $0.097 \pm 0.002      $      \\
$\alpha$ (rad)          &   $4.103 \pm 0.005      $      \\
$\rho$ ($10^{-3}$)      &   $40.79 \pm 0.82       $      \\
\hline
\end{tabular*}
\end{table}

In Table~\ref{table:two}, we list the lensing parameters found from the modeling. We found a 
unique solution with the binary lensing parameters of $(s, q)\sim (1.59, 0.097)$. The inset in 
the lower panel of Figure~\ref{fig:one} shows the lens-system configuration, in which the source 
trajectory (line with an arrow) with respect to the caustic (red closed figure) is presented. For 
the coordinate center of the configuration, we adopt the barycenter for a close binary and the 
effective lens position, defined by \citet{Stefano1996} and \citet{An2002}, for a wide binary. 
In the case of OGLE-2016-BLG-0890, the coordinates are centered at the effective position of 
the lower-mass component, $M_2$, and thus the primary, $M_1$, is located on the right side. 
The topology of the binary lens corresponds to the intermediate regime forming a single merged 
resonant caustic.  To be noted among the lensing parameters is that the normalized source radius, 
$\rho=(40.79\pm 0.82) \times 10^{-3}$, is substantially larger than the typical value of order 
$10^{-3}$ for events involved with main-sequence source stars, and thus the source is likely to 
be a giant star.  The source crossed the caustic lying around the lower-mass lens component with 
a source trajectory angle of $\alpha\sim 55^\circ$. After the first caustic crossing, the source 
swept one fold of the caustic, and this caused the deviation of the light curve from a U-shape 
pattern in the region between the two caustic bumps. It was found that the weak bump at 
${\rm HJD}^\prime \sim 7535$ was produced by the source approach to the primary of the binary 
lens.  The higher-order lensing parameters could not be securely constrained due to the short 
time scale, $\te \sim 15$~days, of the event.

\begin{figure}[t]
\includegraphics[width=\columnwidth]{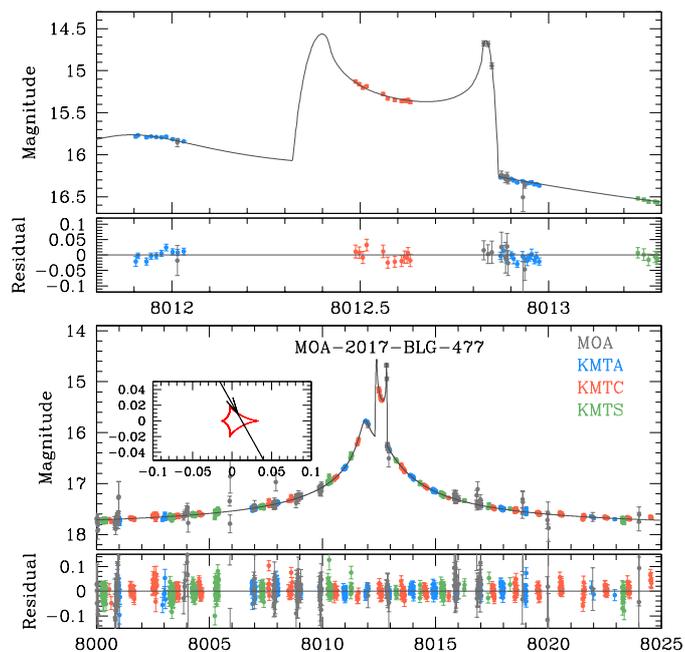}
\caption{
Light curve and model of MOA-2017-BLG-477. Notations and arrangement of the panels 
are same as those in Fig.~\ref{fig:one}.
}
\label{fig:two}
\end{figure}

\begin{table}[t]
\small
\caption{Models of MOA-2017-BLG-477\label{table:three}}
\begin{tabular*}{\columnwidth}{@{\extracolsep{\fill}}lcccc}
\hline\hline
\multicolumn{1}{c}{Parameter}    &
\multicolumn{1}{c}{Close}        &
\multicolumn{1}{c}{Wide }       \\
\hline
$\chi^2$/dof            &   $9264.6/9249          $    &   $9231.2/9249          $     \\
$t_0$ (HJD$^\prime$)    &   $8012.543 \pm 0.003   $    &   $8012.527 \pm 0.003   $     \\
$u_0$                   &   $0.0109 \pm 0.0002    $    &   $0.0117 \pm 0.0002    $     \\
$\te$ (days)            &   $25.10 \pm 0.37       $    &   $26.87 \pm 0.36       $     \\
$s$                     &   $0.355 \pm 0.006      $    &   $3.204 \pm 0.066      $     \\
$q$                     &   $0.097 \pm 0.005      $    &   $0.115 \pm 0.007      $     \\
$\alpha$ (rad)          &   $4.224 \pm 0.007      $    &   $4.206 \pm 0.006      $     \\
$\rho$ ($10^{-3}$)      &   $0.88 \pm 0.06        $    &   $0.80 \pm 0.06        $     \\
\hline
\end{tabular*}
\end{table}

\subsection{MOA-2017-BLG-477}\label{sec:three-two}

The source of the lensing event MOA-2017-BLG-477, with a baseline magnitude of $I_{\rm base}=18.02$, 
lies at the equatorial coordinates $({\rm RA}, {\rm DEC}) _{\rm J2000}=($18:05:50.00, -27:04:38.50), 
which correspond to the Galactic coordinates $(l, b)=(3^\circ\hskip-2pt .854, -2^\circ\hskip-2pt .918)$.  
The MOA group first found the event on 2017 September 15 (${\rm HJD}^\prime\sim 8011$), which was one 
day before the event reached its peak. The event was also observed by the KMTNet group using its three 
telescopes and it was designated as KMT-2017-BLG-1757.

Figure~\ref{fig:two} shows the light curve constructed with the combined MOA and KMTNet data. It 
is found that the peak region of the light curve exhibits three bumps at $t_1\sim 8012.0$, $t_2\sim 
8012.5$, and $t_3\sim 8012.8$. The last peak, covered by the MOA data, appears to be a caustic-crossing 
bump from its shape. Considering that caustic bumps appear in pairs, the bump at $t_2$ would correspond 
to the U-shape region between a pair of caustic bumps arising after an uncovered bump generated by the 
source star's caustic entrance.  The light-curve profile of the bump at $t_1$ is rather smooth, suggesting 
that this bump would be produced by a source approach to a cusp of a caustic.

From modeling, we found two sets of solutions, one in the close-binary regime and the other
in the wide-binary regime.  The two solutions result from the well-known close--wide degeneracy,
which was first mentioned by \citet{Griest1998} and later its origin was investigated by 
\citet{Dominik1999} and \citet{An2005}.  The binary lensing parameters are $(s, q)_{\rm close} 
\sim (0.36, 0.10)$ for the close solution, and $(s, q)_{\rm wide}\sim (3.20, 0.11)$ for the wide 
solution. The full lensing parameters of the individual solutions are presented in 
Table~\ref{table:three}.  It was found that the wide solution yields a better fit to the data 
than the close solution by $\Delta\chi^2=33.4$, which corresponds to $\sqrt{\Delta\chi^2}=5.8\sigma$ 
difference assuming a gaussian error distribution.

The lens-system configuration corresponding to the wide solution is shown in the inset of the
lower panel in Figure~\ref{fig:two}.  Because the lens is in the wide-binary regime, there are 
two sets of caustics according to this solution, and we present the region around the caustic 
through which the source passed.  As expected, the bump at $t_3$ was produced by the caustic 
exit of the source, and the bump at $t_1$ was generated by the cusp approach of the source. 
According to the model, the source entered the caustic at ${\rm HJD}^\prime \sim 8012.42$, which
could have been covered  by the KMTS data if the sky had not been clouded out.  Fortunately, 
the caustic exit was resolved by the 3 data points acquired from MOA observations, and this 
enables us to measure the normalized source radius of $\rho\sim 0.8\times 10^{-3}$. The event 
time scale, $\te \sim 27$~days, is not long enough for us to securely measure the higher-order 
lensing parameters.

\subsection{OGLE-2017-BLG-0614}\label{sec:three-three}
   
The source star of the event OGLE-2017-BLG-0614, lying at the equatorial coordinates of 
$({\rm RA}, {\rm DEC})_{\rm J2000}=($17:26:08.08, -30:17:46.14) and Galactic coordinates of 
$(l, b)=(-3^\circ\hskip-2pt .430, 2^\circ\hskip-2pt .833)$, is very faint, with a baseline 
magnitude of $I_{\rm base}=20.04$.  The alert of the event was issued on 2017 April 23 
(${\rm HJD}^\prime \sim 7867.4$) by the OGLE group at around the peak time of the light curve. 
There appeared to be a single anomalous point around the peak at ${\rm HJD}^\prime \sim 7863$ 
in the OGLE data, but it was difficult to figure out its nature due to the lack 
of data covering the anomaly. The KMTNet group also found the event, labeled as KMT-2017-BLG-2209, 
from the post-season analysis, and 
found that the peak region was well covered by the data from 
the three KMTNet telescopes.

\begin{figure}[t]
\includegraphics[width=\columnwidth]{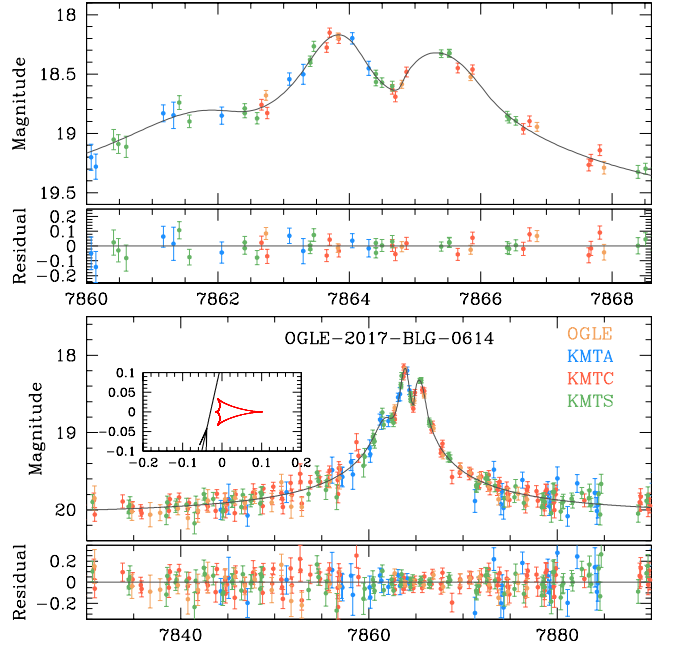}
\caption{
Light curve and model of OGLE-2017-BLG-0614.
}
\label{fig:three}
\end{figure}

The light curve of OGLE-2017-BLG-0614 constructed by combining the OGLE and KMTNet data 
sets is displayed in Figure~\ref{fig:three}. It shows that there exist three smooth bumps 
around the peak region: a weak bump at $t_1\sim 7861.3$ and two relatively strong bumps at 
$t_2\sim 7863.5$ and $t_3\sim 7865.3$. The single anomalous OGLE data point corresponds to 
the peak of the second bump. In general, a caustic crossing produces a sharp spike feature, 
but the feature can be smooth if the source is substantially larger than the caustic. For 
OGLE-2017-BLG-0614, however, the source is likely to be a very faint star, meaning that the 
source is unlikely to be big enough to make caustic-crossing features smooth. This implies 
that all the three observed bumps are likely to be produced by the successive approaches of 
the source to three cusps of a caustic.

\begin{table}[t]
\small
\caption{Models of OGLE-2017-BLG-0614\label{table:four}}
\begin{tabular*}{\columnwidth}{@{\extracolsep{\fill}}lcccc}
\hline\hline
\multicolumn{1}{c}{Parameter}    &
\multicolumn{1}{c}{Close}        &
\multicolumn{1}{c}{Wide }       \\
\hline
$\chi^2$/dof            &   $732.1/724           $    &   $730.4/724           $     \\
$t_0$ (HJD$^\prime$)    &   $7863.991 \pm 0.022  $    &   $7863.991 \pm 0.023  $     \\
$u_0$                   &   $0.028 \pm 0.004     $    &   $0.029 \pm 0.003     $     \\
$\te$ (days)            &   $40.64 \pm 4.70      $    &   $39.82 \pm 4.09      $     \\
$s$                     &   $0.533 \pm 0.016     $    &   $1.842 \pm 0.073     $     \\
$q$                     &   $0.049 \pm 0.006     $    &   $0.050 \pm 0.006     $     \\
$\alpha$ (rad)          &   $1.802 \pm 0.020     $    &   $1.788 \pm 0.021     $     \\
$\rho$ ($10^{-3}$)      &   --                        &   --       \\
\hline
\end{tabular*}
\end{table}

Modeling the light curve yielded two sets of solutions resulting from the close--wide degeneracy, 
with binary parameters of $(s, q)_{\rm close}\sim (0.53, 0.05)$ and $(s, q)_{\rm wide}\sim 
(1.84, 0.05)$ for the close and wide solutions, respectively. The full lensing parameters 
of the two solutions are listed in Table~\ref{table:four}. The degeneracy between the two 
solutions is severe, and the wide model is preferred only by $\Delta\chi^2 =1.7$. The lens 
system configuration for the wide solution is presented in the inset of the lower panel of 
Figure~\ref{fig:three}.  We note that the configuration of the close solution is similar to 
it.  According to the configuration, the bumps were produced by the successive approaches 
of the source to the three cusps of the caustic, as expected from the shapes of the bumps.  
These successive approaches were possible because the three cusps of the caustic lie on one 
side of the primary star due to the small mass ratio, $q\sim 0.05$, between the lens components. 
The strength of the bump varies depending on the combination of the strength of the cusp and 
the separation from the source.  Because none of the bumps were produced by a caustic crossing, 
the normalized source radius could not be measured.  Furthermore, the higher-order lensing 
parameters were difficult to be measured because the precision of the photometric data is not 
high enough to detect the subtle deviations induced by the higher-order effects.

\subsection{KMT-2018-BLG-0357}\label{sec:three-four}

The lensing event KMT-2018-BLG-0357 was observed solely by the KMTNet survey.  The alert on 
the detection of the event was issued on 2018 July 30 (${\rm HJD}^\prime \sim 8330$) with the 
operation of the AlertFinder system system of the KMTNet survey. The source, lying at 
(RA, DEC)$_{\rm J2000}=($17:44:12.20, -33:36:23.18) and $(l, b)=(-4^\circ\hskip-2pt .143, 
2^\circ\hskip-2pt .180)$, has a baseline magnitude of $I_{\rm base}=19.92$.

\begin{figure}[t]
\includegraphics[width=\columnwidth]{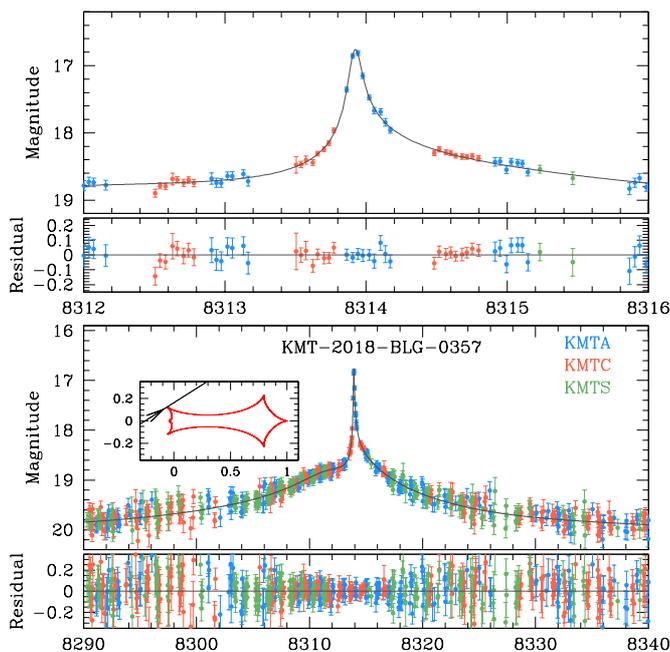}
\caption{
Light curve and model of KMT-2018-BLG-0357.
}
\label{fig:four}
\end{figure}

\begin{table}[t]
\small
\caption{Models of KMT-2018-BLG-0357\label{table:five}}
\begin{tabular*}{\columnwidth}{@{\extracolsep{\fill}}lcccc}
\hline\hline
\multicolumn{1}{c}{Parameter}    &
\multicolumn{1}{c}{Close}        &
\multicolumn{1}{c}{Wide }       \\
\hline
$\chi^2$/dof            &   $1510.5/1483         $    &   $1490.6/1483         $     \\
$t_0$ (HJD$^\prime$)    &   $8313.012 \pm 0.079  $    &   $8313.476 \pm 0.069  $     \\
$u_0$                   &   $0.131 \pm 0.0159    $    &   $0.138 \pm 0.013     $     \\
$\te$ (days)            &   $28.25 \pm 2.27      $    &   $26.70 \pm 2.19      $     \\
$s$                     &   $0.672 \pm 0.012     $    &   $1.461 \pm 0.033     $     \\
$q$                     &   $0.094 \pm 0.009     $    &   $0.111 \pm 0.009     $     \\
$\alpha$ (rad)          &   $2.472 \pm 0.027     $    &   $2.574 \pm 0.028     $     \\
$\rho$ ($10^{-3}$)      &   $1.57 \pm 0.49       $    &   $1.27 \pm 0.46       $     \\
\hline
\end{tabular*}
\end{table}

Figure~\ref{fig:four} shows the light curve of the event constructed with the use of the 
three data sets from the KMTA, KMTC, and KMTS telescopes. It exhibits a strong short-term 
anomaly near the peak at around HJD$^\prime\sim 8313.8$. The central part of the anomaly 
was covered by the KMTA data set, and the peripheral parts on the rising and falling sides 
were covered by the KMTC data set. The anomaly exhibits a typical pattern arising when a 
source approaches or crosses the tip of a caustic cusp.

According to the models, the anomaly was produced by a binary containing a low-mass companion.  
We find two solutions with $(s, q)_{\rm close}\sim (0.67, 0.09)$ and $(s, q)_{\rm wide}\sim 
(1.46, 0.11)$, between which the wide solution is preferred over the close solution by 
$\Delta\chi^2=19.9$. The full lensing parameters of the two solutions are listed in 
Table~\ref{table:five}.

Considering that a short-term anomaly can be produced by a binary companion to the source
\citep{Gaudi1998}, we additionally conducted a binary-source modeling. From this, it is 
found that the binary-source interpretation of the anomaly is excluded with a strong 
statistical confidence of $\Delta\chi^2=480$.

In the inset of the lower panel in Figure~\ref{fig:four}, we present the lens-system 
configuration corresponding to the wide solution.  It shows that the binary lens is in the 
intermediate regime with a single merged caustic, and the anomaly was produced by the source 
crossing over the tip of the off-axis cusp that is closer to the heavier lens component.
The caustic crossing allows us to measure the normalized source radius of $\rho \sim 1.3
\times 10^{-3}$, although its uncertainty is fairly big.

\begin{figure}[t]
\includegraphics[width=\columnwidth]{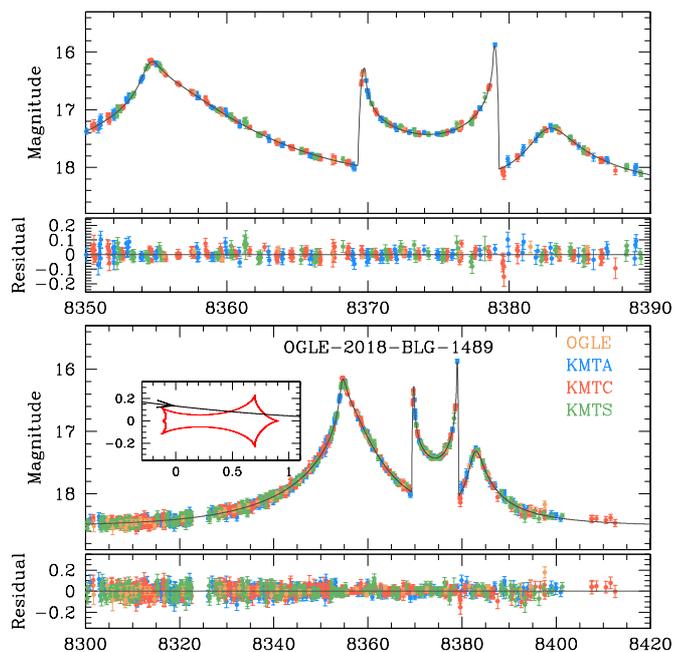}
\caption{
Light curve and model of OGLE-2018-BLG-1489.
}
\label{fig:five}
\end{figure}

\begin{table}[t]
\small
\caption{Models of OGLE-2018-BLG-1489\label{table:six}}
\begin{tabular*}{\columnwidth}{@{\extracolsep{\fill}}lcccc}
\hline\hline
\multicolumn{1}{c}{Parameter}    &
\multicolumn{1}{c}{Standard}        &
\multicolumn{1}{c}{Higher order}       \\
\hline
$\chi^2$/dof                  &   $2094.2/2111         $    &   $2039.6/2108         $     \\
$t_0$ (HJD$^\prime$)          &   $8357.396 \pm 0.014  $    &   $8359.221 \pm 0.047  $      \\
$u_0$                         &   $0.139 \pm 0.001     $    &   $   0.132 \pm 0.001  $      \\
$\te$ (days)                  &   $25.99 \pm 0.04      $    &   $  26.48  \pm 0.17   $     \\
$s$                           &   $1.414 \pm 0.001     $    &   $   1.436 \pm 0.006  $      \\
$q$                           &   $0.097 \pm 0.001     $    &   $   0.103 \pm 0.002  $      \\
$\alpha$ (rad)                &   $3.262 \pm 0.002     $    &   $   3.257 \pm 0.004  $      \\
$\rho$ ($10^{-3}$)            &   $3.93 \pm 0.14       $    &   $   3.86  \pm 0.17   $     \\
$\pien$                       &   --                        &   $  -0.059 \pm 0.51   $     \\ 
$\piee$                       &   --                        &   $  -0.023 \pm 0.06   $     \\
$ds/dt$ (yr$^{-1}$)           &   --                        &   $  -0.72  \pm 0.16   $     \\
$d\alpha/dt$  (rad~yr$^{-1}$) &   --                        &   $   0.07  \pm 0.69   $     \\
\hline                                                                                        
\end{tabular*}
\end{table}

\subsection{OGLE-2018-BLG-1489}\label{sec:three-five}
      
The lensing magnification of the event OGLE-2018-BLG-1489 occurred on a source lying at 
(RA, DEC)$_{\rm J2000}=($17:45:46.60, -23:57:43.85), which correspond to 
$(l, b)=(4^\circ\hskip-2pt .267, 2^\circ\hskip-2pt .559)$. The baseline magnitude of the 
source was $I_{\rm base}=18.55$. The event was first found by the OGLE group on 2018 August 
12 (HJD$^\prime \sim 8343.4$) when the source flux was magnified by about 2.5~mag.  The event 
was found independently by the KMTNet group, who labeled the event as KMT-2018-BLG-1534, from 
the post-season investigation of the data obtained during the 2018 season.

The light curve, shown in Figure~\ref{fig:five}, exhibits a complex pattern with four peaks: 
at HJD$^\prime \sim 8354.1$ ($t_1$), 8369.7 ($t_2$), 8378.7 ($t_3$), and 8382.5 ($t_4$). All 
the anomaly features were well delineated by the data from the KMTNet observations conducted 
with a 1-hour cadence using its three telescopes. From the sharp rise and fall of the light 
curve, it appears that the two peaks at $t_2$ and $t_3$ were produced by the caustic crossings 
of the source.  On the other hand, from the smooth rising and declining of the light curve, it 
appears that the two peaks at $t_1$ and $t_4$ were produced by the cusp approaches. The rising 
part of the caustic entrance at $t_2$ was partially resolved by the KMTC data, 
thus allowing the normalized source radius to be measured.

\begin{figure}[t]
\includegraphics[width=\columnwidth]{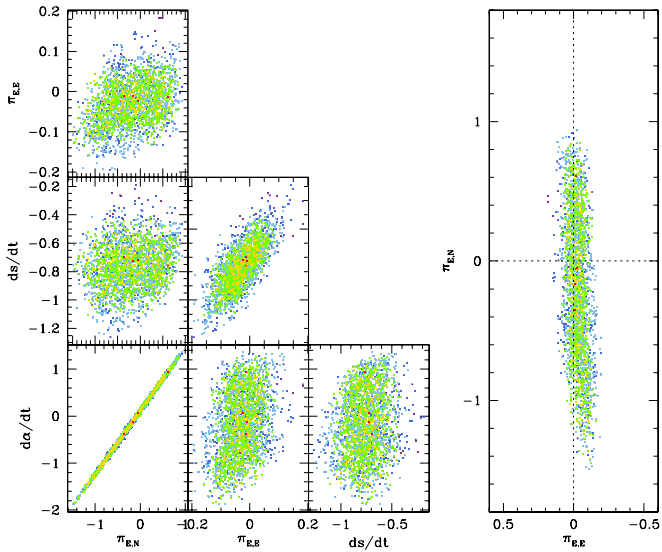}
\caption{
Scatter plot of points in the MCMC chain on the planes of higher-order lensing parameters
of the lensing event OGLE-2018-BLG-1489.  The plot on the $\piee$--$\pien$ parameter plane 
is separately presented in the right panel.  The colors are  set to indicate points with 
$\leq 1\sigma$ (red), $\leq 2\sigma$ (yellow), $\leq 3\sigma$ (green), $\leq 4\sigma$ (cyan), 
and $\leq 5\sigma$ (blue).  
}
\label{fig:six}
\end{figure}

A 2L1S modeling of the light curve yielded a unique solution with binary parameters of $(s, q)
\sim (1.41, 0.1)$. We list the full lensing parameters in Table~\ref{table:six}. In the inset 
of the lower panel in Figure~\ref{fig:five}, we present the configuration of the lens system. 
It shows that the caustic is in the resonant regime, in which a single merged caustic is 
elongated along the binary axis. The source closely approached the upper left cusp at 
$t_1$, entered the caustic at $t_2$, exited the caustic at $t_3$, and then passed by the right 
on-axis cusp of the caustic. The measured value of the normalized source radius is $\rho =
(3.93\pm 0.14)\times 10^{-3}$.

Because the event was continuously covered with a relative good photometric precision, we 
checked whether the higher-order lensing parameters can be constrained.  From an additional 
modeling, it was found that the consideration of the higher-order effects substantially 
improves the model fit by $\Delta\chi^2=54.6$.  The lensing parameters obtained from this 
modeling are listed in Table~\ref{table:six}.  However, it was found that the uncertainties 
of the measured microlens-parallax parameters are large.  In Figure~\ref{fig:six}, we present 
the scatter plots of MCMC points on the planes of higher-order parameters.  From the 
$\piee$--$\pien$ plot, which is separately presented on the right side, it is found that the 
uncertainty of the north component of the parallax vector is substantial.  In the higher-order 
modeling, we impose a restriction that the projected kinetic-to-potential energy ratio is less 
than (KE/PE)$_\perp \leq$ 0.8.  The ratio is computed from the lensing parameters by 
\begin{equation}
\left( {{\rm KE}\over{\rm PE}}\right)_\perp=
{(a_\perp/{\rm AU})\over 8\pi^2(M/M_\odot)}
\left[ 
\left( 
{1\over s} {ds\over dt}
\right)^2 +
\left( {d\alpha\over dt}\right)^2
\right].
\label{eq1}
\end{equation}
Here $a_\perp=\dl\thetae$ is the projected semi-major axis, $M$ and $\dl$ denote the mass 
and distance to the lens, respectively, which are related to the lensing parameters by
\begin{equation}
M={\thetae\over \kappa\pie};\qquad
\dl = {{\rm AU} \over \pie\thetae+ \pi_{\rm S}},
\label{eq2}
\end{equation}
where $\pie=(\pien^2+\piee^2)^{1/2}$, $\kappa=4G/(c^2{\rm AU})$, and $\pi_{\rm S}={\rm AU}/
D_{\rm S}$ \citep{Gould2000}.  It is known that there exists a degeneracy between the parallax 
and orbital effects as discussed in detail by \citet{Skowron2011}.  From the combined facts 
that the region around $(\piee, \pien)= (0, 0)$ is within $2\sigma$ region from the best-fit 
model and that the orbital parameter $|ds/dt|\sim 0.8$, the improvement of the fit relative 
to the standard model is mostly ascribed to the lens-orbital effect rather than the 
microlens-parallax effect. Nevertheless, one component of the microlens parallax, i.e., 
$\piee$, is well constrained.

\begin{figure}[t]
\includegraphics[width=\columnwidth]{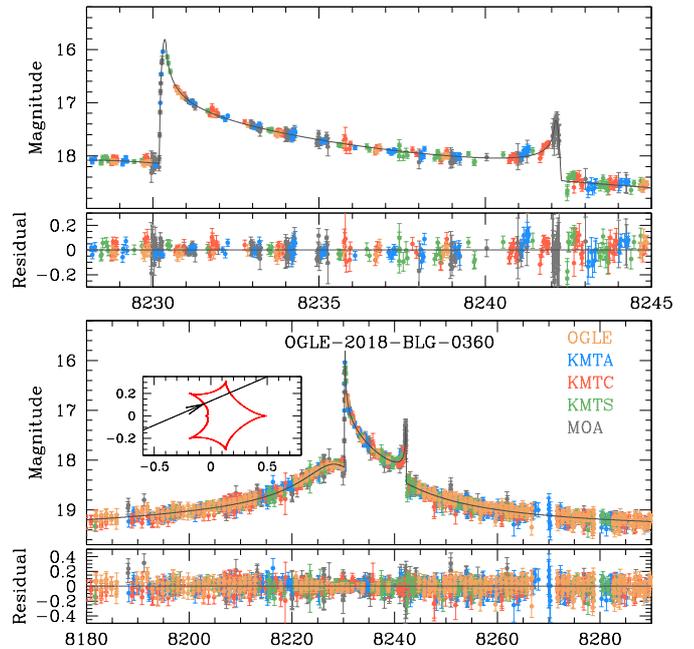}
\caption{
Light curve and model of OGLE-2018-BLG-0360.
}
\label{fig:seven}
\end{figure}

\begin{table}[t]
\small
\caption{Model of OGLE-2018-BLG-0360\label{table:seven}}
\begin{tabular*}{\columnwidth}{@{\extracolsep{\fill}}lcccc}
\hline\hline
\multicolumn{1}{c}{Parameter}    &
\multicolumn{1}{c}{Value }       \\
\hline
$\chi^2$/dof            &   $3576.3/3569         $      \\
$t_0$ (HJD$^\prime$)    &   $8230.867 \pm 0.048  $      \\
$u_0$                   &   $0.122 \pm 0.002     $      \\
$\te$ (days)            &   $47.63 \pm 0.78      $      \\
$s$                     &   $1.056 \pm 0.003     $      \\
$q$                     &   $0.063 \pm 0.002     $      \\
$\alpha$ (rad)          &   $2.732 \pm 0.010     $      \\
$\rho$ ($10^{-3}$)      &   $2.18 \pm 0.04       $      \\
\hline
\end{tabular*}
\end{table}

\begin{table*}[t]
\small
\caption{Properties of source stars\label{table:eight}}
\begin{tabular}{llcccc}
\hline\hline
\multicolumn{1}{c}{Events}    &
\multicolumn{1}{c}{$(V-I, I)_{\rm S}$}    &
\multicolumn{1}{c}{$(V-I, I)_{\rm RGC}$}    &
\multicolumn{1}{c}{$I_{0,{\rm RGC}}$}      &
\multicolumn{1}{c}{$(V-I, I)_{0,{\rm S}}$}    &
\multicolumn{1}{c}{$\theta_*$ ($\mu$as)}       \\
\hline
OGLE-2016-BLG-0890   & $(3.45\pm 0.16, 15.979\pm 0.001)$  &  $(3.09, 16.725)$    &  14.573 &  $(1.42\pm 0.16, 13.827\pm 0.001)$   & $10.35\pm 1.82 $   \\
MOA-2017-BLG-477     & $(1.80\pm 0.01, 19.800\pm 0.005)$  &  $(2.28, 15.448)$    &  14.332 &  $(0.58\pm 0.10, 18.684\pm 0.005)$   & $0.50 \pm 0.04 $  \\
OGLE-2017-BLG-0614   & $(3.37\pm 0.12, 22.242\pm 0.005)$  &  $(3.54, 17.067)$    &  14.596 &  $(0.89\pm 0.12, 19.771\pm 0.005)$   & $0.43 \pm 0.06 $  \\
KMT-2018-BLG-0357    & $(2.59\pm 0.12, 21.040\pm 0.011)$  &  $(2.88, 16.993)$    &  14.372 &  $(0.77\pm 0.12, 18.423\pm 0.011)$   & $0.69 \pm 0.09 $  \\
OGLE-2018-BLG-1489   & $(2.14\pm 0.02, 18.970\pm 0.003)$  &  $(2.47, 16.182)$    &  14.322 &  $(0.73\pm 0.02, 17.110\pm 1.003)$   & $1.36 \pm 0.10 $  \\
OGLE-2018-BLG-0360   & $(2.73\pm 9.13, 20.319\pm 0.008)$  &  $(2.91, 17.238)$    &  14.512 &  $(0.88\pm 0.13, 17.593\pm 0.008)$   & $1.15 \pm 0.30 $  \\
\hline
\end{tabular}
\tablefoot{ $(V-I)_{0,{\rm RGC}}=1.06$  }
\end{table*}

\subsection{OGLE-2018-BLG-0360}\label{sec:three-six}
     
The source star of the lensing event OGLE-2018-BLG-0360 lies at the equatorial and Galactic
coordinates of (RA, DEC)$_{\rm J2000}=($17:52:01.26, -31:08:54.71) and $(l, b)=
(1^\circ\hskip-2pt .183, -2^\circ\hskip-2pt.326)$, respectively.  The baseline magnitude of 
the source is $I_{\rm base}=19.28$. The event was observed by all of the three currently operating 
microlensing surveys.  The OGLE group first detected the event on 2018 March 15 (HJD$^\prime =8193.4$), 
the MOA group, who labeled the event as MOA-2018-BLG-116, found it on 2018-04-22 (HJD$^\prime=
8231.4$), and the KMTNet group identified the event, labeled as KMT-2018-BLG-2014, from the 
post-season investigation of the data.

Figure~\ref{fig:seven} shows the light curve of the event constructed by combining the data 
from the three survey experiments.  It shows a characteristic pattern of a binary-lens event 
with two caustic-crossing spikes, for which the first spike at HJD$^\prime\sim 8230$ was covered 
by the combination of MOA, KMTA, and KMTS data sets, and the second one at HJD$^\prime\sim 8242$ 
was resolved by the MOA data set.  In addition to these spikes, there is a weak bump at 
HJD$^\prime\sim 8228$.

Modeling the light curve yielded a unique solution with binary parameters of $(s, q)\sim (1.06, 
0.06)$, indicating that the event was produced by a binary in an intermediate regime with a 
low-mass companion. We list the full lensing parameters in Table~\ref{table:seven}. The 
normalized source radius estimated from the analysis of the caustic-crossing parts is $\rho=
(2.18\pm 0.04)\times 10^{-3}$. According to the lens-system configuration, presented in the 
inset of the lower panel, the spikes were produced by the source crossings over the two folds 
of the 6-sided resonant caustic that are separated by consecutive off-axis cusps, and the weak 
bump was produced by the approach of the source close to the on-axis cusp near the host.  From 
the modeling considering higher-order effects, it was found that the microlens parallax was 
difficult to be securely measured because of the moderate photometric precision of the data.

\begin{figure}[t]
\includegraphics[width=\columnwidth]{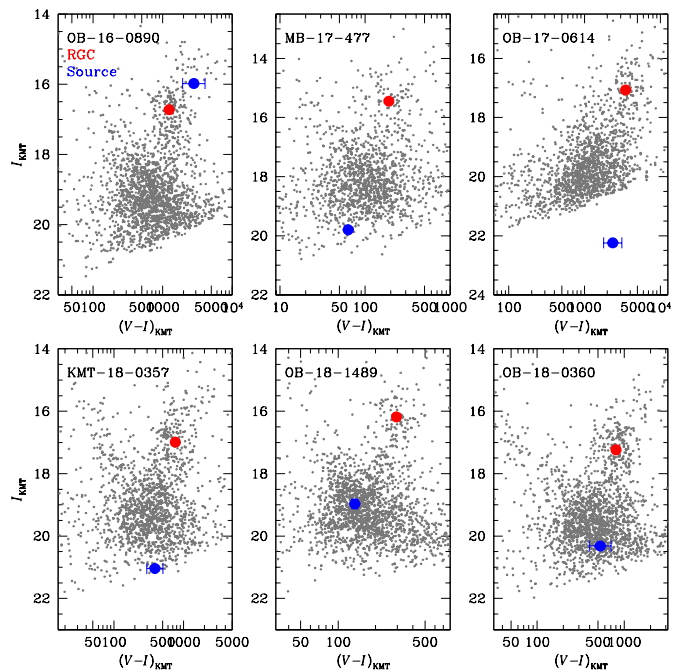}
\caption{
Locations of the source stars (blue filled dots) of the individual events with respect to 
the centroids of red giant clump (red dot) in the instrumental color-magnitude diagrams 
constructed from the pyDIA photometry of KMTNet data.
}
\label{fig:eight}
\end{figure}

\begin{table}[t]
\small
\caption{Einstein radius and proper motion\label{table:nine}}
\begin{tabular*}{\columnwidth}{@{\extracolsep{\fill}}lcccc}
\hline\hline
\multicolumn{1}{c}{Event}                        &
\multicolumn{1}{c}{$\thetae$ (mas)}              &
\multicolumn{1}{c}{$\mu$ (mas~yr$^{-1}$) }       \\
\hline
OGLE-2016-BLG-0890     &  $0.26 \pm 0.05$    &  $6.30 \pm 1.12$     \\
MOA-2017-BLG-477       &  $0.67 \pm 0.06$    &  $9.33 \pm 0.83$     \\
OGLE-2017-BLG-0614     &  --                 &   --                 \\
KMT-2018-BLG-0357      &  $0.57 \pm 0.08$    &  $7.52 \pm 1.05$     \\
OGLE-2018-BLG-1489     &  $0.36 \pm 0.03$    &  $4.89 \pm 0.36$     \\
OGLE-2018-BLG-0360     &  $0.53 \pm 0.08$    &  $4.12 \pm 0.59$     \\
\hline
\end{tabular*}
\end{table}

\section{Source stars and angular Einstein radii}\label{sec:four}

For 5 out of 6 analyzed lensing events, the normalized source radii were securely measured 
from the deviations of the light curves affected by finite-source effects. In this section, 
we estimate the angular Einstein radii for these events. The value of $\thetae$ is estimated 
from the measured $\rho$ value by 
\begin{equation}
\thetae = {\theta_*\over \rho},
\label{eq3} 
\end{equation}
where the angular source radius $\theta_*$ is estimated from the color and brightness of the 
source.  Although the Einstein radius cannot be measured for OGLE-2017-BLG-0614 because $\rho$ 
value could not measured, we estimate $\theta_*$ for the full characterization of the event.

The angular source radius of each event was estimated from the $V-I$ color and $I$-band
magnitude. For the estimation of the reddening and extinction-corrected (dereddened) color 
and magnitude, $(V-I, I)_{\rm 0,S}$, from instrumental values, we  apply the method of 
\citet{Yoo2004}. In this method, the centroid of red giant clump (RGC), with known dereddened 
values of $(V-I, I)_{\rm 0,RGC}$, in the color-magnitude diagram (CMD) is used as a reference 
to convert instrumental color and magnitude $(V-I, I)$ into $(V-I, I)_{\rm 0,S}$, that is,
\begin{equation}
(V-I, I)_{\rm 0,S} = (V-I, I)_{\rm 0,RGC} + [(V-I, I)_{\rm S} - (V-I, I)_{\rm RGC}]. 
\label{eq4} 
\end{equation}
Here $(V-I, I)_{\rm S}$ and $(V-I, I)_{\rm RGC}$ denote the instrumental colors and magnitudes 
of the source and RGC, respectively, and thus the term in the bracket on the right side of 
Equation~(\ref{eq4}) indicates the offsets in color and magnitude of the source from the RGC 
centroid in the instrumental CMD.  For this conversion, the dereddened color, $(V-I)_{\rm 0,RGC}
=1.06$, and magnitude of the RGC centroid were adopted from \citet{Bensby2013} and 
\citet{Nataf2013}, respectively.

Figure~\ref{fig:eight} shows the instrumental CMDs of stars lying near the source stars of 
the individual events constructed with the use of the photometry data processed using the pyDIA 
\citep{Albrow2017} reductions of the KMTC data. In each diagram, we mark the locations of the 
source and RGC, indicated by red and blue filled dots, respectively. The $I$- and $V$-band magnitudes 
of each source were measured from the regression of the photometry data in the individual passbands 
processed using the same pyDIA code with the variation of the lensing magnification. In 
Table~\ref{table:eight}, we summarize the values of $(V-I, I)_{\rm S}$, $(V-I, I)_{\rm RGC}$, 
$I_{\rm RGC}$, and $(V-I, I)_{\rm 0,S}$ for the individual events. According to the estimated 
values of $(V-I, I)_{\rm 0,S}$, it is found that the source of OGLE-2016-BLG-0890 is a K-type 
giant, and those of the other events are main-sequence stars with spectral types ranging from 
G to K.  We then converted $V-I$ into $V-K$ using the color-color relations of \citet{Bessell1988} 
and then derived $\theta_*$ from the \citet{Kervella2004} relation between $(V-K, V)$ and $\theta_*$.  
The estimated source radii of the individual events are listed in the last column of 
Table~\ref{table:eight}.

With the measured source radii, the angular Einstein radii of the events were estimated using 
the relation in Equation~(\ref{eq3}). With the measured event time scale, the relative proper 
motion between the lens ans source was estimated by
\begin{equation}
\mu = {\thetae \over \te}.
\label{eq5}
\end{equation}
In Table~\ref{table:nine}, we list the estimated values of $\thetae$ and $\mu$ of the individual 
events.  In the cases of the events MOA-2017-BLG-477 and KMT-2018-BLG-0357, for which two models 
were presented, we present $\thetae$ and $\mu$ values estimated from the wide models, which 
yield better fits over the corresponding close solutions with significant confidence levels 
of $\Delta\chi^2=33.4$ and $19.9$, respectively.  It was found that the Einstein radii of 
the events lie in the range of [0.26--0.67]~mas, and the proper motions are in the range of 
[4.1--9.3]~mas~yr$^{-1}$.

\begin{table*}[t]
\small
\caption{Physical lens parameters \label{table:ten}}
\begin{tabular}{lllllrrrrr}
\hline\hline
\multicolumn{1}{c}{Events}            &
\multicolumn{1}{c}{$M_1$}             &
\multicolumn{1}{c}{$M_2$}             &
\multicolumn{1}{c}{$\dl$}             &
\multicolumn{1}{c}{$a_\perp$}         &
\multicolumn{1}{c}{$P_{\rm BD}$}      &
\multicolumn{1}{c}{$P_{\rm star}$}    &
\multicolumn{1}{c}{$P_{\rm planet}$}  &
\multicolumn{1}{c}{$P_{\rm disk}$}    &
\multicolumn{1}{c}{$P_{\rm bulge}$}  \\
\multicolumn{1}{c}{}                  &
\multicolumn{1}{c}{($M_\odot$)}       &
\multicolumn{1}{c}{($M_\odot$)}       &
\multicolumn{1}{c}{(kpc)}             &
\multicolumn{1}{c}{(AU)}              &
\multicolumn{1}{c}{(\%)}              &
\multicolumn{1}{c}{(\%)}              &
\multicolumn{1}{c}{(\%)}              &
\multicolumn{1}{c}{(\%)}              &
\multicolumn{1}{c}{(\%)}              \\ 
\hline
OGLE-2016-BLG-0890           & $0.40^{+0.35}_{-0.20} $   &  $0.038^{+0.033}_{-0.020}$  &  $7.4^{+0.8}_{-1.1} $  & $3.5^{+0.4}_{-0.4}  $  & 86   &  5  &  9  &  1  &  99 \\ [0.8ex]
MOA-2017-BLG-477 (wide)      & $0.74^{+0.33}_{-0.36} $   &  $0.085^{+0.040}_{-0.041}$  &  $5.0^{+1.0}_{-1.4} $  & $10.5^{+2.0}_{-3.0} $  & 61   & 37  &  2  & 62  &  38 \\ [0.8ex]
OGLE-2017-BLG-0614 (close)   & $0.64^{+0.38}_{-0.37} $   &  $0.031^{+0.019}_{-0.018}$  &  $6.4^{+1.6}_{-2.4} $  & $1.8^{+0.4}_{-0.7}  $  & 81   &  1  & 18  & 53  &  47 \\ [0.8ex]
\hskip89pt         (wide)    &  --                       &  $0.032^{+0.019}_{-0.018}$  &   --                   & $6.0^{+1.5}_{-2.3}  $  & --   & --  & --  & --  &     \\ [0.8ex]
KMT-2018-BLG-0357 (wide)     & $0.68^{+0.34}_{-0.34} $   &  $0.075^{+0.035}_{-0.036}$  &  $6.2^{+1.2}_{-1.7} $  & $5.0^{+1.0}_{-1.3}  $  & 69   & 28  &  3  & 62  &  38 \\ [0.8ex]
OGLE-2018-BLG-1489           & $0.48^{+0.32}_{-0.19} $   &  $0.050^{+0.031}_{-0.020}$  &  $6.5^{+0.9}_{-1.0} $  & $3.5^{+0.5}_{-0.5}  $  & 89   &  9  &  2  & 39  &  61 \\ [0.8ex]
OGLE-2018-BLG-0360           & $0.72^{+0.31}_{-0.35} $   &  $0.045^{+0.019}_{-0.021}$  &  $6.4^{+0.1}_{-1.6} $  & $3.6^{+0.6}_{-0.9}  $  & 90   &  2  &  8  & 49  &  51 \\ [0.8ex]
\hline
\end{tabular}
\end{table*}

\section{Physical lens properties}\label{sec:five}

In addition to the basic observable of $\te$, unique determinations of the lens mass and distance
require one to additionally measure two extra observables of $\thetae$ and $\pie$ by the relation 
given in Equation~(\ref{fig:two}).  For all events except OGLE-2018-BLG-1489, the microlens-parallax 
could not be measured, and even for OGLE-2018-BLG-1489, the uncertainty of the measured $\pie$ is 
very big, as shown in Figure~\ref{fig:six}.  As a result, it is difficult to uniquely determine $M$ 
and $\dl$ from the relations in Equation~(\ref{eq2}).  Although the $\pie$ constraint is either 
unavailable or weak, one can still constrain the physical lens parameters using the other observables 
with the use of a Galactic model defining the distributions of mass density, motion, and mass function 
of Galactic objects. For these constraints, we conduct Bayesian analyses of the individual events.

The Bayesian analyses were done according to the following procedure. In the first step, we generated 
a large number ($6\times 10^6$) of artificial lensing events. For the individual events, their 
physical parameters of the lens mass $M$, distances to the lens $\dl$ and source $D_{\rm S}$, 
and lens-source transverse velocity $v_\perp$ were assigned from the Monte Carlo simulation 
conducted with the use of a Galactic model.  In the simulation, we adopted the \citet{Jung2021} 
Galactic model.  For the mass density distribution, the Galactic model uses the \citet{Robin2003} 
disk model and \citet{Han2003} bulge model. For the kinematic distribution of disk objects, the 
model uses the modified version of the \citet{Han1995} model, in which the original version based 
on the double-exponential disk model was modified to reconcile it with the the \citet{Robin2003} 
density distribution. For the kinematic distribution of bulge objects, the model was constructed 
based on the proper motions of stars in the Gaia catalog \citep{Gaia2016, Gaia2018}. For the mass 
functions of bulge and disk populations, the Galactic model adopted the initial mass function and 
the present-day mass function of \citet{Chabrier2003}, respectively. See \citet{Jung2021} for 
details of the Galactic model.

In the second step, we constructed the posterior distributions of the physical lens parameters. 
For this, we first computed event time scales and Einstein radii of the artificial events produced 
by the simulation as $t_{{\rm E},i}=\dl\theta_{{\rm E},i}/v_\perp$ and $\theta_{{\rm E},i}=
(\kappa M \pi_{\rm rel})^{1/2}$, respectively. Then, the Bayesian posteriors of each lensing 
event are constructed by assigning a weight for each artificial event as $w_i=\exp(-\chi^2/2)$. 
Here $\chi^2=(t_{{\rm E},i}-\te)^2/[\sigma(\te)]^2+(\theta_{{\rm E},i}-\thetae)^2/[\sigma(\thetae)]^2$, 
where $[\te, \sigma(\te)]$ and $[\thetae, \sigma(\thetae)]$ are the measured values of $\te$ and 
$\thetae$ and their uncertainties, respectively. We note that only $\te$ is measured for 
OGLE-2017-BLG-0614, while both observables of $\te$ and $\thetae$ are measured for the other events. 
In the case of OGLE-2018-BLG-1489, for which the model with higher-order effects is better than the 
standard model with $\Delta\chi^2=54.6$, we impose an additional constraint of $\pivec_{\rm E}$ given 
by the covariance matrix of the parallax ellipse presented in Figure~\ref{fig:six}. In the cases of 
the events MOA-2017-BLG-477 and KMT-2018-BLG-0357, for which the wide solutions are favored over the 
close solutions with significant statistical confidence, we conduct Bayesian analyses for the wide 
solutions. In the case of OGLE-2017-0614, for which the degeneracy is very severe, we carry out 
Bayesian analysis for both the close and wide solutions.  The source star of OGLE-2016-BLG-0890 
is bright, and thus it is registered in the Gaia catalog.  In this case, we use the additional 
constraint of the source proper motion, $(\mu_E, \mu_N)=(-5.170 \pm 0.180, -11.385 \pm 0.123)~{\rm 
mas~yr^{-1}}$, in the Bayesian analysis.

\begin{figure}[t]
\includegraphics[width=\columnwidth]{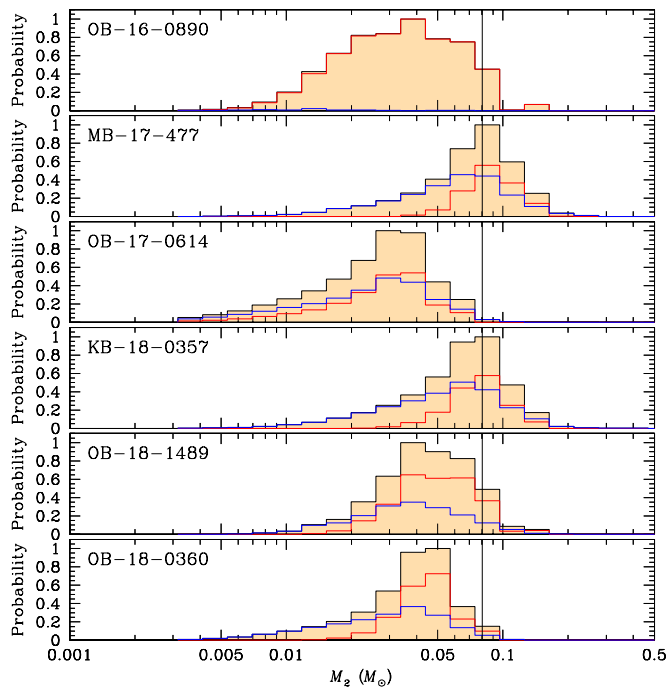}
\caption{
Bayesian posterior for the companion masses of the binary lenses. The solid vertical line in 
each panel represents the mass boundary between a brown dwarf and a star. The blue and red 
curves represent the contributions by the disk and bulge lens populations, respectively, and 
the black curve is the sum of the contributions.
}
\label{fig:nine}
\end{figure}

\begin{figure}[t]
\includegraphics[width=\columnwidth]{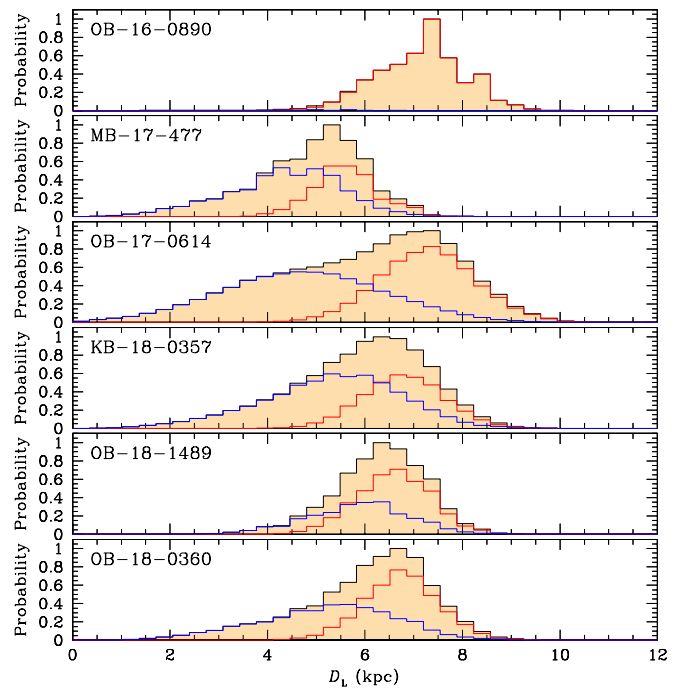}
\caption{
Bayesian posterior for the distances to the lens systems. Notations are same as those in Fig.~\ref{fig:nine}.
}
\label{fig:ten}
\end{figure}

Figures~\ref{fig:nine} and \ref{fig:ten} show the posterior distributions of the companion lens 
mass and the distance to the lens systems for the individual lensing events, respectively. For 
each distribution, the blue and red curves represent the contributions by the disk and bulge 
lens populations, respectively, and the black curve is sum of the contributions from the two 
lens populations.  In Table~\ref{table:ten}, we summarize the values of $M_1$, $M_2$, $\dl$, 
and $a_\perp$, for which the median values are listed as representative values, the uncertainties 
are estimated as the 16\% and 84\% of the posterior distributions, and $a_\perp=s\dl \thetae$ 
represents the projected separation between the binary lens components.  Also presented in the 
table are the probabilities for the individual events that the lens companions are BD ($P_{\rm BD}$), 
star ($P_{\rm star}$), or planet ($P_{\rm planet}$) and disk ($P_{\rm disk}$) or bulge ($P_{\rm bulge}$) 
members.  According to the posteriors of $M_2$, the probabilities for the lens companions of the events 
OGLE-2016-BLG-0890, OGLE-2017-BLG-0614, OGLE-2018-BLG-1489, and OGLE-2018-BLG-0360 to be in the BD mass 
regime of [0.012 -- 0.08]~$M_\odot$ are very high with $P_{\rm BD}> 80\%$.  For MOA-2017-BLG-477 and 
KMT-2018-BLG-0357, the probabilities are $P_{\rm BD}=61\%$ and 69\%, respectively, and one cannot 
completely rule out the possibility that the companions of the lenses are very low-mass stars. 
In our Bayesian analyses, we assume that the primary and companion follow the same mass function.
If the number of companions in the BD regime declines compared to the mass function of the primary, 
that is, brown-dwarf desert, for example, \citet{Grether2006}, the BD probability $P_{\rm BD}$ 
would be less than the presented probabilities.

\section{Summary and conclusion}\label{sec:six}

We investigated the microlensing survey data collected during the 2016--2018 seasons with the aim 
of finding microlensing binaries containing BD companions.  In order to sort out BD-companion 
binary-lens events, we conducted modeling of all lensing events detected during the seasons with 
lensing light curves exhibiting anomaly features that were likely to be produced by binary lenses, 
and then applied the criterion that the companion-to-primary mass ratio was less than 0.1.  From 
this procedure, we found 6 candidate BD binary events including OGLE-2016-BLG-0890, MOA-2017-BLG-477, 
OGLE-2017-BLG-0614, KMT-2018-BLG-0357, OGLE-2018-BLG-1489, and OGLE-2018-BLG-0360, for which analyses 
had not been presented before.

For the identified candidate events, we conducted detailed modeling using optimized photometry 
data and checked possible degenerate interpretations caused by various types of degeneracies.  
We also checked the feasibility of detecting higher-order effects.  We presented the solutions 
of the individual events and the corresponding lens-system configurations.

According to the estimated masses of the binary companions, we found that the probabilities for 
the lens companions of the events OGLE-2016-BLG-0890, OGLE-2017-BLG-0614, OGLE-2018-BLG-1489, and 
OGLE-2018-BLG-0360 to be in the BD mass regime were very high.  For the companions of the events 
MOA-2017-BLG-477 and KMT-2018-BLG-0357, it was found that the median masses were near the star-BD 
boundary, and thus the possibilities that the companions of the lenses were very low-mass stars could 
not be completely ruled out.  

\begin{acknowledgements}
Work by C.H. was supported by the grants of National Research Foundation of Korea 
(2020R1A4A2002885 and 2019R1A2C2085965).
This research has made use of the KMTNet system operated by the Korea Astronomy and Space 
Science Institute (KASI) and the data were obtained at three host sites of CTIO in Chile, 
SAAO in South Africa, and SSO in Australia.
The MOA project is supported by JSPS KAKENHI
Grant Number JSPS24253004, JSPS26247023, JSPS23340064, JSPS15H00781,
JP16H06287, and JP17H02871.
J.C.Y.  acknowledges support from NSF Grant No. AST-2108414.
W.Z. and H.Y. acknowledge support by the National Science Foundation of
China (Grant No. 12133005). 
C.R. was supported by the Research fellowship of the Alexander von Humboldt Foundation.
\end{acknowledgements}


\begin{thebibliography}{}
\bibitem[Albrow et al.(2009)]{Albrow2009} Albrow, M., Horne, K., Bramich, D.~M., et al.\ 2009, \mnras, 397, 2099
\bibitem[An(2005)]{An2005} An, J. H. 2005, \mnras, 356, 1409
\bibitem[Albrow(2017)]{Albrow2017} Albrow, M.\ 2017, MichaelDAlbrow/pyDIA: Initial Release on Github,Versionv1.0.0, Zenodo, doi:10.5281/zenodo.268049
\bibitem[Albrow et al.(2000)]{Albrow2000}  Albrow, M. D., Beaulieu, J.-P., Caldwell, J. A. R., et al. 2000, \apj, 534, 894
\bibitem[Alcock etal.(1996)]{Alcock1996}  Alcock, C., Allsman, R. A., Axelrod, T. S., et al.\ 1996, \apj, 461, 84
\bibitem[An(2005)]{An2005}  An, J. H. 2005, \mnras, 356, 1409
\bibitem[An \& Han(2002)]{An2002}  An, J. H., \& Han, C. 2002, \apj, 573, 351
\bibitem[Aubourg et al.(1995)]{Aubourg1995}  Aubourg, E., Bareyre, P., Brehin, S., et al. 1995, \aap, 301, 1
\bibitem[Bennett \& Rhie(1996)]{Bennett1996} Bennett, D. P., \& Rhie, S. H. 1996, \apj, 472, 660
\bibitem[Bensby et al.(2013)]{Bensby2013}  Bensby, T. Yee, J.C., Feltzing, S. et al. 2013, \aap, 549, A147
\bibitem[Bessell \& Brett(1988)]{Bessell1988}  Bessell, M.~S., \& Brett, J.~M. 1988, \pasp, 100, 1134
\bibitem[Bond et al.(2001)]{Bond2001}  Bond, I. A., Abe, F., Dodd, R. J., et al. 2001, \mnras, 327, 868
\bibitem[Cassan(2008)]{Cassan2008}  Cassan, A. 2008, \aap, 491, 587
\bibitem[Chabrier(2003)]{Chabrier2003}  Chabrier, G. 2003, \pasp, 115, 763
\bibitem[Chung et al.(2019)]{Chung2019}  Chung, S.-J. Gould, A., Skowron, J., et al. 2019, \apj, 871, 179
\bibitem[Di Stefano \& Mao(1996)]{Stefano1996}  Di Stefano, R., \& Mao, S. 1996, \apj, 457, 93
\bibitem[Dominik(1999)]{Dominik1999}  Dominik, M. 1999, \aap, 349, 108
\bibitem[Erdl \& Schneider(1993)]{Erdl1993}  Erdl, H., \& Schneider, P. 1993, \aap, 268, 453
\bibitem[Gaia Collaboration(2016)]{Gaia2016}  Gaia Collaboration, Prusti, T., de Bruijne, J. H. J., et al. 2016, \aap, 595, A1
\bibitem[Gaia Collaboration(2018)]{Gaia2018}  Gaia Collaboration, Brown, A. G. A., Vallenari, A., et al. 2018, \aap, 616, A1
\bibitem[Gaudi(1998)]{Gaudi1998} Gaudi, B. S. 1998, \apj, 506, 533
\bibitem[Gould(1992)]{Gould1992b}  Gould, A. 1992, \apj, 392, 442
\bibitem[Gould(2000)]{Gould2000}  Gould, A. 2000, \apj, 542, 785
\bibitem[Gould \& Loeb(1992)]{Gould1992a} Gould, A. \& Loeb, A. 1992, \apj, 396, 104
\bibitem[Gould et al.(2022)]{Gould2022} Gould, A., Jung, Y.K., Hwang, K.-H., et al., 2022, JKAS, submitted, arXiv:2204.03269
\bibitem[Grether \& Lineweaver(2006)]{Grether2006} Grether, D., \& Lineweaver, C. H. 2006, \apj, 640, 1051
\bibitem[Griest \& Safizadeh(1998)]{Griest1998}  Griest, K., \& Safizadeh, N. 1998, \apj, 500, 37
\bibitem[Han et al.(2019)]{Han2019} Han, C., Bond, I. A., Udalski, A., et al. 2019, \apj, 876, 81 
\bibitem[Han \& Gould(2003)]{Han2003}  Han, C., \& Gould, A. 2003, \apj, 592, 172
\bibitem[Han \& Gould(1995)]{Han1995}  Han, C., \& Gould, A. 1995, \apj, 447, 53
\bibitem[Han et al.(2022)]{Han2022}  Han, C., Gould, A., Bond, I. A., et al. 2022, \aap. 662, A70
\bibitem[Han et al.(2021a)]{Han2021a}   Han, C., Lee, C.-U., Ryu, Y.-H., et al. 2021a, \aap, 649, A91 
\bibitem[Han et al.(2020a)]{Han2020a}  Han, C., Lee, C.-U., Udalski, A., et al. 2020a, \aj, 159, 134
\bibitem[Han et al.(2020b]{Han2020b}  Han, C., Kim, D., Udalski, A., et al. 2020b, \aj, 160, 64
\bibitem[Han et al.(2020c)]{Han2020c}  Han, C., Udalski, A., Kim, D. et al. 2020c, \aap, 642,A110
\bibitem[Han et al.(2021b)]{Han2021b}  Han, C., Udalski, A., Kim, D. et al. 2021b, \aap, 655, A21
\bibitem[Herald et al.(2022)]{Herald2022} Herald, A., Udalski, A., Bozza, A., et al. 2022, \aap, in press,  arXiv:2203.04034
\bibitem[Jung et al.(2021)]{Jung2021}  Jung, Y. K., Han, C., Udalski, A., et al. 2021, \aj, 161, 293
\bibitem[Jung et al.(2018)]{Jung2018} Jung, Y. K., Udalski, A., Gould, A., et al. 2018, \aj, 155, 219
\bibitem[Jung et al.(2017)]{Jung2017} Jung, Y. K., Udalski, A., Yee, J. C., et al. 2017, \aj, 153,129 
\bibitem[Kervella et al.(2004)]{Kervella2004} Kervella, P., Th\'evenin, F., Di Folco, E., \& S\'egransan, D.\ 2004, \aap, 426, 29
\bibitem[Kim et al.(2018b)]{Kim2018b}  Kim, D.-J., Kim, H.-W., Hwang, K.-H., et al. 2018a, \aj, 155, 76
\bibitem[Kim et al.(2018a)]{Kim2018a}  Kim, H.-W., Hwang, K.-H., Shvartzvald, Y., et al. 2018b, arXiv:1806.07545
\bibitem[Kim et al.(2016)]{Kim2016}  Kim, S.-L., Lee, C.-U., Park, B.-G., et al. 2016, JKAS, 49, 37
\bibitem[Mao \& Paczy\'nski(1991)]{Mao1991}  Mao, S., \& Paczy\'nski, Bohdan 1991, \apj, 374, L37
\bibitem[Miyazaki et al.(2018)]{Miyazaki2018} Miyazaki, S., Sumi, T., Bennett, D. P., et al. 2018, \aj, 156, 136
\bibitem[Nataf et al.(2013)]{Nataf2013}  Nataf, D. M., Gould, A., Fouqu\'e, P. et al. 2013, \apj, 769, 88
\bibitem[Paczy\'nski(1986)]{Paczynski1986}  Paczy\'nski, B. 1986, \apj, 304, 1
\bibitem[Robin et al.(2003)]{Robin2003}  Robin, A. C., Reyl\'e, C., Derri\'ere, S., \& Picaud, S. 2003, \aap, 409, 523
\bibitem[Shvartzvald et al.(2019)]{Shvartzvald2019}  Shvartzvald, Y., Yee, J. C., Skowron, J., et al. 2019, \aj, 157, 106
\bibitem[Skowron et al.(2011)]{Skowron2011} Skowron, J., Udalski, A., Gould, A., et al. 2011, \apj, 738, 87
\bibitem[Udalski(2003)]{Udalski2003}  Udalski, A. 2003, Acta Astron., 53, 291
\bibitem[Udalski et al.(2015)]{Udalski2015}  Udalski, A., Szyma\'nski, M. K., \& Szyma\'nski, G. 2015, Acta Astron., 65, 1
\bibitem[Udalski et al.(1993)]{Udalski1993}  Udalski, A., Szyma\'nski, M., Ka{\l}u\.zny, J., Kubiak, M., Krzemi\'nski, W., 
Mateo, M., Preston, G. W., \& Paczy\'nski, B. 1993, Acta Astron., 43, 289 
\bibitem[Wo\'zniak(2000)]{Wozniak2000} Wo\'zniak, P. R. 2000, Acta Astron., 50, 421 
\bibitem[Yee et al.(2012)]{Yee2012} Yee, J. C., Shvartzvald, Y., Gal-Yam, A., et al.\ 2012, \apj, 755, 102
\bibitem[Yoo et al.(2004)]{Yoo2004}  Yoo, J., DePoy, D.L., Gal-Yam, A. et al. 2004, \apj, 603, 139
\vspace*{\fill}
\end{thebibliography}
\end{document}